\documentclass[12pt]{article}

\usepackage{amsmath,amstext,amssymb,amsthm}
\usepackage{xspace}
\usepackage{url}
\usepackage{hyperref}
\usepackage{graphicx,epsfig}
\usepackage{fullpage}

\newtheorem{theorem}{Theorem}[section]

\newcommand{\rddots}{\mathinner{
    \mkern2mu\raise1pt\hbox{.}
    \mkern2mu\raise4pt\hbox{.}
    \mkern1mu\raise7pt\vbox{\kern7pt\hbox{.}}
    \mkern1mu}}

\newcommand{\QQ}{{\mathbb Q}}
\newcommand{\ZZ}{{\mathbb Z}}
\newcommand{\RR}{{\mathbb R}}
\newcommand{\FF}{{\mathbb F}}

\makeatletter
\renewcommand\section{\@startsection {section}{1}{\z@}%
                                   {-3.5ex \@plus -1ex \@minus -.2ex}%
                                   {2.3ex \@plus.2ex}%
                                   {\normalfont\large\bfseries}}
\renewcommand\subsection{\@startsection{subsection}{2}{\z@}%
                                     {-3.25ex\@plus -1ex \@minus -.2ex}%
                                     {1.5ex \@plus .2ex}%
                                     {\normalfont\normalsize\bfseries}}

\makeatother

\begin{document}

\begin{center}
{\bf CERTIFICATION OF THE QR FACTOR R, \\[0.1cm] AND OF LATTICE BASIS  REDUCEDNESS}\\[0.3cm]
Gilles Villard\\[0.3cm]
{\footnotesize
Laboratoire LIP (CNRS, ENSL, INRIA, UCBL)\\ \'Ecole Normale
Sup\'erieure de Lyon, France\\
\url{http://perso.ens-lyon.fr/gilles.villard}\\
} 
\end{center}

\renewcommand{\thefootnote}{\fnsymbol{footnote}}

\footnotetext{\!\!\!\!\!\!\!\!\!\!This 
material is based on work supported
in part by the French  National Research Agency, ANR
Gecko. \\[-0.3cm]
~~\\
{{\em LIP Research Report RR2007-03, \'Ecole Normale
    Sup\'erieure de Lyon ---  January, 2007.}}

}

{\abstract  
\noindent
\hspace*{0.6cm}
Given a lattice basis of $n$ vectors in $\ZZ ^n$, 
we propose an algorithm using
$12n^3+O(n^2)$ floating point
operations for checking whether  
the basis is LLL-reduced.
If the basis is reduced then the algorithm will hopefully answer
``yes''. If the basis is not reduced, or if the precision used is not
sufficient with respect to $n$, and to the numerical properties of
the basis, the algorithm will answer ``failed''.
Hence a positive answer is a rigorous certificate. 
For implementing the certificate itself, we propose a floating point
algorithm
for computing (certified) error bounds for the entries of the $R$ factor of the
$QR$ matrix factorization. This algorithm takes into account all possible
approximation and rounding errors.\\
\hspace*{0.6cm}The cost $12n^3+O(n^2)$ of the certificate is
only six times more than the cost of  
numerical algorithms for computing the $QR$ factorization itself, and the
certificate may be implemented using matrix library routines only.
We report experiments 
that show that for a reduced basis of 
adequate dimension and quality the certificate succeeds, and
establish the effectiveness of the certificate. This effectiveness
is applied for certifying the output of fastest existing floating point
heuristics of LLL reduction, without slowing down the whole process.
}


\section{Introduction}

Our motivation is to develop a  
certificate for lattice basis reducedness that may be 
used in cooperation with---possibly non certified---numerical
reduction heuristics such as those described
in~\cite[Ch.\,II-3]{Ste05} and~\cite{NgSt06}.
The two main constraints are speed and effectiveness. Indeed, 
 the certificate has to be fast enough for not
slowing down the whole process, and the answer should be 
relevant (``yes'')  on a large class of inputs such as those
successfully treated by the heuristic. Hence   
our general concern is somehow the compromize between speed 
and proven accuracy. 
The certificate will be
introduced later below. It relies on error
bounds for the $R$ factor of the $QR$
factorization of a matrix that we discuss first. \\[-0.2cm]

\noindent
{\bf Bounding errors for the factor $R$.}
Let $A$ be an $n \times n$ invertible integer matrix. 
The $QR$ factorization (see for instance \cite[Ch.\,19]{Hig02})  
of $A$ is a factorization $A=QR$ in which the factor $R \in \RR^{n
  \times n}$ is an upper triangular matrix, and the factor $Q\in \RR^{n
  \times n}$ is orthogonal ($Q^TQ$=I). We take the unique
factorization such that the diagonal entries of $R$ are positive.  
Let $\FF$ denote a set of floating point numbers such that  
the arithmetic operations in  $\FF$
satisfy the 
IEEE~754 arithmetic standard~\cite{IEEE754}. 
Assume that an approximate floating point and upper triangular  factor $\widetilde{R}\in \FF^{n
  \times n}$ 
is given. 
In Section~\ref{sec-errorR}
we propose 
an algorithm for computing a componentwise error bound for
$|\widetilde{R}-R|$ using operations   
in $\FF$ only. For a matrix 
$A=(a_{i,j})$, 
$|A|$ denotes $(|a_{i,j}|)$.
Our error bound for 
$|\widetilde{R}-R|$ is given by  
a matrix $H\in \FF^{n
  \times n}$ with positive entries such   
that (see~(\ref{eq-errorH}) on page~\pageref{eq-errorH}):
\begin{equation} \label{eq-defE}
|\widetilde{R}-R| \leq H |\widetilde{R}|.
\end{equation}
Since floating point numbers are rational numbers, 
when $\widetilde{R}$ and $E$ are known, (\ref{eq-defE}) provides  
a rigourous mathematical bound for the error with respect to the unkown matrix $R$. 

For understanding the behaviour
of the error bounding algorithm better, 
we recall in Section~\ref{sec-errorQR} some existing  
numerical pertubation analyses for the $QR$ factorization. 
The necessary background material may be found in Higham's
book~\cite{Hig02}. Then in Sections~\ref{sec-sun} and~\ref{sec-tow},
we give the mathematical foundations of our approach. We 
focus 
on the componentwise bounds of~\cite{Sun92} that allow
us to derive an algorithm 
based on the principles of verification (self-validating)
methods. On the latter methods we refer to the rich surveys of
Rump~\cite{Rum03,Rum05},
see also the short discussion in
Section~\ref{sec-verif}.
As numerical experiments of Section~\ref{subsec-comput} will demonstrate,
the error bounding algorithm is effective in practice. 
Its cost is only $5$ times more than a numerical
$QR$ factorization, we mean $10n^3+O(n^2)$ operations in~$\FF$.
For efficiency, the error bounds are themselves calculated using
floating point operations, nevertheless, they take into account all
possible numerical and rounding errors. The reducedness
certificate will require $2n^3+O(n^2)$ additional operations.
Most of the $12n^3$ operations actually correspond to the evaluation of
matrix expressions. An efficient implementation may thus rely on 
fast matrix routines such as the BLAS~\cite{Blas90}. 

At a given precision, the error bounding algorithm  
provides relevant bounds for input matrices with appropriate 
numerical properties. 
In particular, the dimension and related condition numbers  
should be considered in relation with the precision (see Section~\ref{subsec-comput}). 
However, the power of the verification
approach~\cite{Rum03,Rum05}
is to be effective on many inputs for which the numerical
approach itself is effective---here the numerical $QR$ factorization.
For example, we report experiments using $64$~bits floating point numbers, 
and $\widetilde{R}$ computed by the modified 
Gram-Schmidt orthogonalization (see~\cite[Alg.\,19.12]{Hig02}).
On integer matrices of dimension $n=1500$ with 
condition number around $10^5$, we certify 
that the relative error on the entries of 
$\widetilde{R}$ 
has order as small as $10^{-6}$ or $10^{-5}$, 
with only $10^{-10}$ or $10^{-9}$ on the diagonal.
We refer here to the diagonal entries since they play a key role for
instance in the LLL Lov\'asz test (see~(\ref{eq:deflovasz})).
For large condition numbers (with respect to double precision), say $10^{12}$, 
and $n=200$, the algorithm may typically certify relative
errors in $10^{-1}$, and  $10^{-4}$ on the diagonal. \\[-0.2cm]

\noindent
{\bf The LLL-reducedness certificate.}
The effectiveness of the error bound on $|\widetilde{R}-R|$ allows us to
address the second topic of the paper. 
To an $n\times n$ integer matrix $A$ we associate the Euclidean 
lattice ${\mathcal L}$ generated by the columns  
$(a_j)$ of $A$ (for definitions and on algorithmic aspects
of lattices 
we refer for instance to~\cite{Coh95}). 
From $(a_j)$, the LLL algorithm computes
a reduced basis~\cite{LLL82}, where the reduction is defined via the Gram-Schmidt
orthogonalization of $a_1, a_2, \ldots, a_n \in \ZZ ^n$.
The Gram-Schmidt orthogonalization determines the associated orthogonal basis 
$a ^*_1, a ^*_2, \ldots, a ^*_n \in \QQ ^n$ by induction,
together with factors $\mu _{ij}$, using
$a_i ^* = a_i - \sum _{j=1}^{i-1}\mu _{ij}a_j^*$,
and
$\mu _{ij} = \langle a_i, a_j^* \rangle / \|a_j^*\|^2 _2$, $1
\leq j < i$.
Vectors $a_1, a_2, \ldots, a_n$ are said proper
for $\eta \geq 1/2$ if their Gram-Schmidt
orthogonalization satisfies 
\begin{equation} \label{eq:defproper}
|\mu_{ij}|\leq \eta,~1
\leq j < i \leq n.
\end{equation}
In general one
considers $\eta = 1/2$.
The basis $a_1, a_2, \ldots, a_n$
of ${\mathcal L}$ is called LLL-reduced with factors $\delta$ and
$\eta$
if the vectors are proper, and if they satisfy the Lov\'asz
conditions:
\begin{equation} \label{eq:deflovasz}
(\delta - \mu _{i+1,i} ^2) \|a_i^*\|^2 _2 \leq \|a_{i+1}^*\|^2 _2,
~1\leq i \leq n-1,
\end{equation}
with $1/4 < \delta \leq 1$ and $1/2 \leq \eta < \sqrt{\delta}$.
If $A=QR$ is the $QR$ factorization of $A$ then we have
\begin{equation} \label{eq:QRGS}
\left\{ \begin{array}{l}
\|a_i^*\| _2 = r_{ii},~1 \leq i \leq n,\\
\mu _{ij} = r_{ji}/r_{jj},~1
\leq j < i \leq n.
\end{array}
\right.
\end{equation}
We see from~(\ref{eq:QRGS}) that if an approximation $\widetilde{R}$
of $R$ with error bounds on its entries are known,  
then (depending on the quality of the bounds) it may be possible
to check whether~(\ref{eq:defproper}) and~(\ref{eq:deflovasz})
are satisfied. 
All the above draws the reducedness
certificate that we propose in~Section~\ref{sec-reduced}.
We also fix a set $\FF$ of floating point numbers, and perform
operations in $\FF$ only. 
For certifying the reducedness of the column basis associated to $A$
the certificate works in three steps:
\begin{description}
\item {\sc i}: Numerical computation of a $R$ factor $\widetilde{R}$ such that 
  $A\approx \widetilde{Q}\widetilde{R}$; 
\item {\sc ii}: Certified computation of $F \in \FF ^{n \times n}$
  such that $|\widetilde{R}-R| \leq F$ (see (\ref{eq-defE}));
\item {\sc iii}: Certified check of properness (\ref{eq:defproper})
  and Lov\'asz conditions (\ref{eq:deflovasz}). 
\end{description}

Following the principles of verification algorithms~\cite{Rum05},
Step~{\sc i} is purely approximation, and we propose an implementation of 
Steps~{\sc ii} and~{\sc iii}
that is independent of the factorization algorithm used for
computing  $\widetilde{R}$. 
For taking into account all possible numerical and rounding errors,
Steps~{\sc ii} and~{\sc iii} use certified computing
techniques (see Section~\ref{subsec:certiftech}). We rely on the fact that the arithmetic
operations $+, -, \times, \div, \sqrt{~}$ in $\FF$ are according to
the  IEEE~754 standard. We especially use explicit changes of
rounding mode for certified bounds. 

Verification algorithms are a powerful
alternative between numerical  and computer
algebra algorithms, they somehow illustrate the boundary between the
two fields. 
The reducedness certificate we propose illustrates a cooperation of purely
numerical computation  with a certified approach based on  the
IEEE~754 
standard, in order to provide a computer
algebra answer. 
Our progress in linear algebra is in the line of previous works on 
error bounds for linear
systems~\cite{Rum94,OiRu02,RuOg07}, on 
certifying the sign of the determinant~\cite{Pan01,KaVi04}, 
on verifying positive definiteness~\cite{Rum06}, or on eigenvalues~\cite{May94,Rum01}.
Our contribution is to establish the effectiveness of
componentwise bounds for a whole matrix,
propose a corresponding certified algorithm using fast
verification techniques, and derive and test with experiments a certificate 
for the LLL reducedness application.\\[-0.2cm]

\noindent 
{\bf Absolute value and matrix norms.} We already considered above the absolute value of a
matrix $A=(a_{ij})$ defined by  $(|a_{ij}|)$. We write $|A| \leq
|B|$ if $|a_{ij}| \leq |b_{ij}|$. It is possible to check that 
if $A=BC$ then $|A| \leq
|B||C|$. We will use several matrix norms
(see~\cite[Ch.\,6]{Hig02}) such as the Frobenius norm
$\|\cdot\|_{F}$ or the $2$-norm $\|\cdot\|_2$.
We will also especially use the infinity norm
$\|\cdot\|_{\infty}=\max_{1\leq i \leq n} \sum _{j=1}^n |a_{ij}|$.
For $A=BC$ we have $\|A\|_{\infty} \leq
\|B\|_{\infty}\|C\|_{\infty}$, and if $h=\|A\|_{\infty}$ then 
$|A| \leq H$ with $h_{ij}=h$.\\[-0.2cm]

\noindent 
{\bf Condition numbers.} For a nonsingular matrix $A$, the matrix
condition number is defined by  
$\kappa _p (A) = \|A\|_p  \|A ^{-1}\|_p$ with $p=2, F$ or $\infty$~\cite[Th.\,6.4]{Hig02}.
With the infinity norm we will also use the Bauer-Skeel condition number
$\text{cond}(A)=
\||A ^{-1}||A|\|_{\infty} \leq \kappa _{\infty}(A)$~\cite[\S\,7.2]{Hig02}.


\section{Error bounds computation and verification algorithms} \label{sec-verif}

In linear algebra, few things are known
about the complexity of computing certified and effective error bounds. 
The problem is somewhere between the one of computing approximate
solutions, and the one of computing multi-precision or exact 
 solutions. A main result in~\cite{DDM01} 
shows  that 
the problem of computing 
a certified estimation of $\|A ^{-1}\|$ (for a consistent matrix
norm) is as difficult as testing whether the product of two matrices
is zero. Hence if we consider $O(n ^3)$ operations for
multiplying two matrices of dimension $n$, 
a deterministic error bound---based on a condition number bound---would
cost $O(n ^3)$. 
The use of randomization may lead to 
error estimations in  $O(n ^2)$ operations, we refer
to~\cite[Chap.\,15]{Hig02}
and references therein, and to the fact that the matrix product
could be verified in  $O(n ^2)$ operations~\cite{Fre79}.  
We did not investigate the randomization possibilities yet.

Verification methods have been developped in~\cite{Rum94,OiRu02} for
computing certified error bounds 
for linear system solution. In~\cite{OiRu02} the error bound
(normwise) is computed in twice the time of
numerical 
Gaussian elimination. 
In the same spirit,  a verification approach using $O(n^3)$ 
floating point operations is proposed in~\cite{Pan01} for the
sign of the determinant (see~\cite{KaVi04} for a survey on
this topic). Note that computing the sign of the determinant corresponds to knowing
the determinant with a relative error less than $1$.
Our error bounding algorithm for $R$ will also use $O(n^3)$ floating
point operations. 
The verification approach~\cite{Rum03,Rum05}  
gives an effective alternative to interval arithmetic whose
exponential overestimation of the error would not be appropriate for
our problem~\cite[\S10.7]{Rum05}. The general strategy for
calculating an error bound 
is first to establish a result whose assertion is  a mathematical expression for the
bound (see Theorem~\ref{theo-error}), then design an
algorithm that verifies the assumptions for  
the latter assertion, and computes a certified evaluation of the bound
(see Section~\ref{sec-tow}).



\section{Perturbation analyses and bounds for the $QR$
  factorization} \label{sec-errorQR}


A finite precision computation of the $QR$ factorization of $A$ leads to
an approximate factor~$\widetilde{R}$. The errors in $\widetilde{R}$ with
respect to $R$ are called the {\em forward errors} (absolute or
relative). The matrix $\widetilde{R}$ is not the factor of the $QR$ factorization of $A$,
however, it is seen as the $QR$ factor 
of a perturbed matrix $\widetilde{A}=A+E$,
where $E$ is called the {\em backward error.}
The choice of $\widetilde{A}$ is
non unique, and one refers for instance for the smallest error norm.
The link between backward and forward error is made using
the condition number of the problem, hence for us the
condition number for the problem of computing  $R$.  
The (relative) {\em condition number} of the problem---under some
class of perturbations---measures the 
relative change in the output for a relative change in the input. 
In this context, a useful tool  
for estimating the accuracy
of the solution to a problem, is the rule of thumb~\cite[p.\,9]{Hig02}:
\begin{equation} \label{thumb}
\text{forward error} 
\begin{array}{c}~\\[-0.36cm]<\\[-0.28cm] \sim\end{array}
\text{condition number} \times \text{backward error}.
\end{equation}
We survey below some more precise
instantiations of~(\ref{thumb}) for the $QR$ 
factorization. 
Known results are, in general, approximate inequalities (first order
results), but could be extended   
for giving strict bounds on the forward error. 
The rule of thumb therefore gives a first possible direction 
for deriving an error bounding algorithm for $|\widetilde{R}-R|$ (the
forward absolute error).
However, most of corresponding bounds rely
on matrix norms, and may thus overestimate  
the actual componentwise error in most cases. 

We will investigate an alternative direction in Section~\ref{sec-sun}.
Rather than on the rule of thumb, our error bounding 
algorithm will be based on the componentwise bounds of Sun~\cite{Sun92}. 
This will lead to an algorithm that seems to be naturally more
effective than a matrix norm approach for our problem.
Another advantage of using Sun's results  is to remain in the spirit of the
verification methods. In particular,  we will
see that the error bounding algorithm is oblivious of the algorithm that is
used for computing the approximate factor $\widetilde{R}$. Our bound computation
may be appended to any numerical $QR$ algorithm, and does
not rely on backward error bounds that would be have been needed for
using (\ref{thumb}). An approximate $\widetilde{Q}$ in not
orthogonal in
general, the backward error problem is to know  for which matrix $\widetilde{A}$ close to $A$, there
exists an orthogonal $\widehat{Q}$ such that  $\widetilde{A}=\widehat{Q} \widetilde{R}$?  
Backward error bounds are known for specific $QR$ algorithms such as Householder
or Gram-Schmidt ones (see Theorems~19.4 and~19.13 in \cite{Hig02}), but
may not be available in the general case. We will circumvent
the need of the backward error in Section~\ref{sec-sun} using the correspondence
between the $QR$ factorization of $A$, and the Cholesky
factorization $R^T R$ of $A ^T A$.   \\[-0.2cm]

\noindent
{\bf Sensitivity of the $QR$ factorization.}
The condition number of the problem of computing $R$ (the ``rate of
change'' of $R$) in the $QR$
factorization may be defined theoretically for given classes of
perturbations, but it is non trivial  to derive expressions
of the condition number that can be used  in practice. Nevertheless, various formulae
are proposed in the literature providing quantities that can be thought
as a condition number for $R$, we refer for instance to~\cite{ChPa01}.
These quantities may be very effective in practice in a matrix norm setting.

Let $A=QR$  and $A \approx \widetilde{A}
+E = \widehat{Q} \widetilde{R}$ be $QR$ factorizations. As already noticed, 
for a floating point factorization $A \approx \widetilde{Q}\widetilde{R}$, in general 
we have $\widehat{Q} \neq \widetilde{Q}$ since $\widetilde{Q}$ is not orthogonal.
Let $\widetilde{R} = R+F$.
For a sufficiently small backward error $E$, consider the normwise
relative error $\epsilon = \|E\|_F / \|A\|_2 =  \|\widetilde{A}-A\|_F / \|A\|_2$. Then
Sun's~\cite[Rem.\,3.5]{Sun91} perturbation bounds (see
also~\cite{Ste93}) give
\begin{equation} \label{eq-sun91}
{\|\widetilde{R}-R\|_F}/{\|R\|_2} \leq \sqrt{2} \kappa _2 (A)
\epsilon + O(\epsilon ^2). 
\end{equation}
An improved bound is given by Zha~[Theorem\,2.1]\cite{Zha93} (see
also~\cite[\S\,5]{ChPa01} and~\cite[\S19.9]{Hig02}) under a componentwise model of
perturbation that we simplify here.   
Let $|\widetilde{A}-A|=|E|=\epsilon |A|$, then for sufficiently small 
$\epsilon$ we have:
\begin{equation} \label{eq-zha}
{\|\widetilde{R}-R\|_{\infty}}/{\|R\|_{\infty}} \leq c_n
\text{cond}(R^{-1}) \epsilon + O(\epsilon ^2) 
\end{equation}
where $c_n$ is a constant depending on $n$.
Hence the Bauer-Skeel condition number of $R^{-1}$ can be considered 
as a condition number for the problem of calculating $R$.
This indicates that 
one may potentially loose significant digits (in the result)
linearly with respect to the increase of $\log
\text{cond}(R^{-1})$. 
This typical behaviour is illustrated by Figure~3.1
where we have computed $QR$ factorizations of random matrices 
(of {\tt randsvd} type~\cite[Ch.\,28]{Hig02}). The algorithm
used is the Modified Gram-Schmidt
algorithm~\cite[Algo.\,19.12]{Hig02}.
\begin{center}
\includegraphics[scale=0.5]{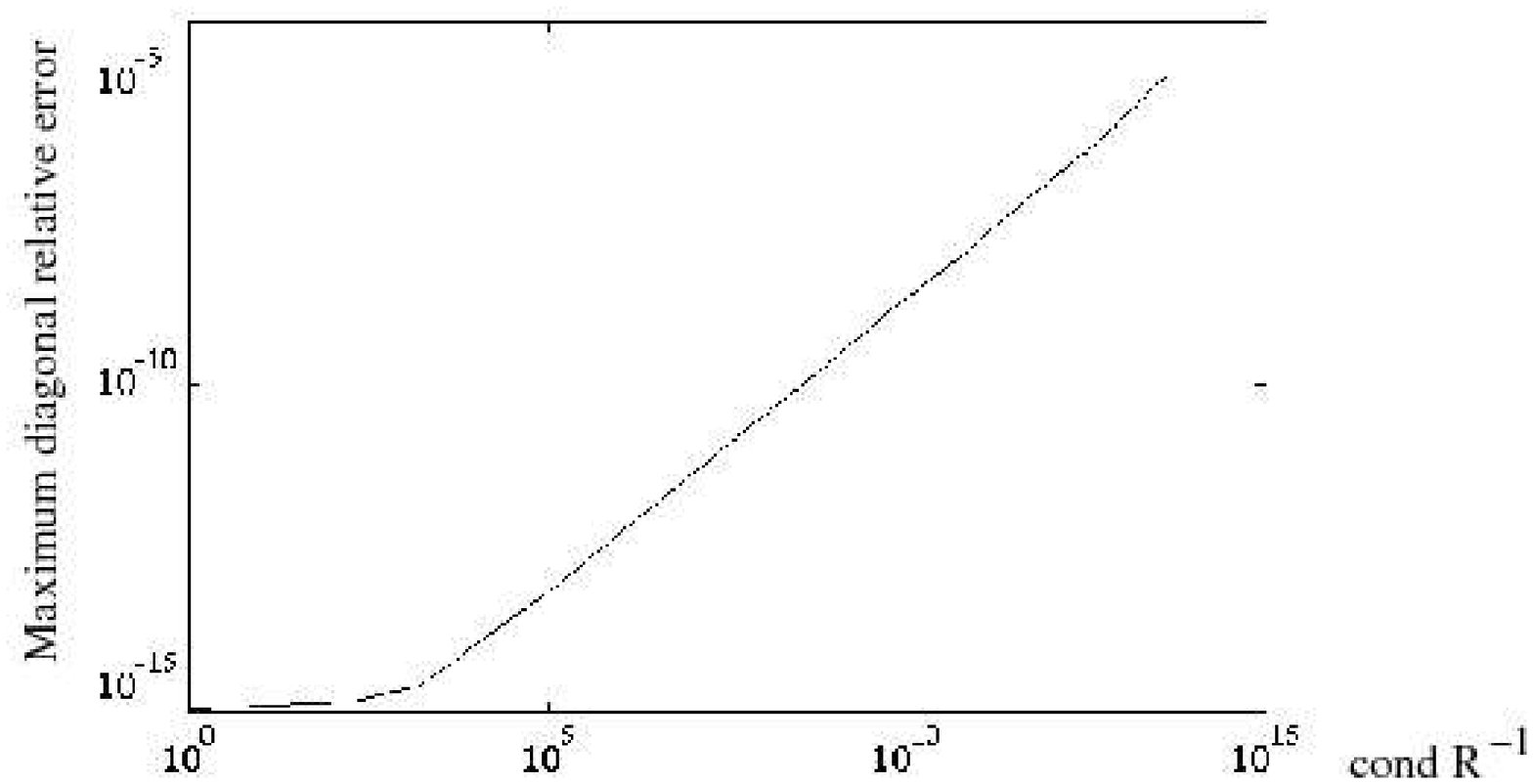} 
{\footnotesize
~\\
\begin{minipage}{13.4cm}\begin{center}
{\footnotesize Figure 3.1: Maximum
  relative diagonal error in $R$ (Modified Gram-Schmidt algorithm) with
  respect to cond($R^{-1}$)
for random matrices $A$ ($n=200$).}\end{center}
\end{minipage}
}
\end{center}Identities~(\ref{eq-sun91}) and~(\ref{eq-zha})
provide first order estimations of the errors. They are essential
for an idea 
of the normwise loss of accuracy. 
Nevertheless, the loss of accuracy on 
individual entries  (needed for the reducedness certificate)
 may not be deduced from these identities.  
Consider for instance the case of Figure~3.1 where the ratios of the $r_{ij}$ 
may be as large as $10^{11}$. The normwise bound of~(\ref{eq-zha}),
that involves the max row sum $\|R\|_{\infty}$, cannot provide relevant informations
for every $|\tilde{r}_{ij}-r_{ij}|$. 
Note also that the 
loss of accuracy would certainly be amplified by the implementation of the 
error estimation itself in finite precision (Figure~3.1
is a mathematical representation of the error). Normwise 
bounds much sharper than~(\ref{eq-sun91}) and~(\ref{eq-zha})
may be found, especially in~\cite{CPS97,ChPa01}, 
it remains to know how well the corresponding proposed estimations approximate the
true condition number~\cite[\S10]{ChPa01}. 
It would also be interesting to investigate how the new techniques 
of~\cite{CPS97,ChPa01} could lead to practical componentwise
bounds. 


\section{Strict componentwise bounds for the $R$ factor} \label{sec-sun}

We now present the mathematical view and justification of the error
bounding algorithm of Section~\ref{sec-errorR}.
Given $A \in \RR ^{n \times n}$ invertible, 
and an upper triangular matrix $\widetilde{R} \in \RR ^{n \times n}$, 
the problem is to bound $|\widetilde{R}-R|$ where $R$ is the unknown 
$QR$ factor of $A$. In practice  we will have $A,\widetilde{R} \in \FF ^{n \times n}$.

\subsection{$QR$ and Cholesky factorization}

The strict componentwise analysis of Sun~\cite[\S\,4]{Sun92}
for $QR$ uses the matrix $\widetilde{A}$ such that 
$\widetilde{A}=\widehat{Q}\widetilde{R}$ is a $QR$ factorization.
Note that, because of the loss of orthogonality,
 if $\widetilde{Q}$ is a numerical approximation of $Q$ then 
$\widetilde{A}$ is not in general the 
matrix $\widetilde{Q}\widetilde{R}$. 
Informations on $\widetilde{A}$ may be available by taking into
account the algorithm that has produced $\widetilde{R}$. 
We refer for instance to~\cite[Eq.\,(5.8)]{ChPa01}
and~\cite[\S19.9\,]{Hig02}
where properties of Householder transformations are used for
bounding the backward error.
This is not sufficient for our problem since we are given only $A$ and
$\widetilde{R}$, and since one of our goal is to be oblivious of the method
used for $\widetilde{R}$.

For not relying on $\widetilde{A}$, we propose to rather resort to
Sun's study of 
the Cholesky factorization~\cite[\S\,4]{Sun92}. 
If $B \in \RR ^{n \times n}$ is symmetric positive definite, then there is a unique upper
triangular $R \in \RR ^{n \times n}$
with positive diagonal entries,  such that $B=R^TR$. This
factorization is called the Cholesky factorization~\cite[Th.\,10.1]{Hig02}.
It holds that $A=QR$ is a $QR$ factorization if and only if 
$B=A ^T A =R^T R$ is a Cholesky factorization.
It may not be a good idea to use the
Cholesky factorization for computing $R$ numerically. The condition
number 
of the problem may indeed increase too much, especially $\kappa _2 (A
^T A) = (\kappa _2 (A)) ^2$. For avoiding this drawback,  
our point is to implement the 
reduceness certificate of Section~\ref{sec-reduced} using  $QR$
for computing $\widetilde{R}$, and to use
the Cholesky point of view only for computing the error bound.

\subsection{The bound on $|\widetilde{R}-R|$}

For a matrix $A \in \RR ^{n \times n}$, 
the spectral radius 
$\rho (A)$ is the maximum of the eigenvalue modules. 
We denote by $\text{\rm triu} (A)$ the upper triangular 
part of $A$, we mean that $\text{\rm triu} (A) = (t_{ij})$ with 
$t_{ij}=a_{ij}$ if $i\leq j$, and $t_{ij}=0$ otherwise.
The following Theorem is~\cite[Th.\,2.1]{Sun92}.

\begin{theorem} \label{theo-errorsun}
For $B, \widetilde{B} \in  \RR ^{n \times n}$ symmetric positive
definite matrices, let $R$ and $\widetilde{R}$ be the 
Cholesky factors of  $B$ and $\widetilde{B}$. Let 
$E= \widetilde{B}-B$,
and 
\begin{equation} \label{eq-defG}
G = |\widetilde{R}^{-T} E  \widetilde{R}^{-1}|.
\end{equation}
Then if 
$\rho (G) < 1$
we have 
\begin{equation} \label{eq-errorH}
|\widetilde{R}-R| \leq \text{\rm triu} (G (I-G)^{-1}) |\widetilde{R}|.
\end{equation}
\end{theorem}
Inequality~(\ref{eq-errorH}) is what we announced
with~(\ref{eq-defE}).
Let us apply Theorem~\ref{theo-errorsun} 
with $B=A ^TA$  and
$\widetilde{B}=\widetilde{A}^T \widetilde{A}$.
Using $\widetilde{A} = \widehat{Q}\widetilde{R}$ and 
$\widehat{Q}^T\widehat{Q}=I$,
we get from~(\ref{eq-defG}):
$$\begin{array}{ll}
G  &= |\widetilde{R}^{-T} E  \widetilde{R}^{-1}|
= |\widetilde{R}^{-T} (\widetilde{B}-B) 
\widetilde{R}^{-1}|
= |\widetilde{R}^{-T} (\widetilde{A}^T\widetilde{A} - A ^T A) 
\widetilde{R}^{-1}|\\
&= |\widetilde{R}^{-T} \widetilde{A}^T\widetilde{A}
\widetilde{R}^{-1}
- \widetilde{R}^{-T} A ^T A 
\widetilde{R}^{-1}| = 
|\widehat{Q}^{T} \widehat{Q}  
- \widetilde{R}^{-T} A ^T A 
\widetilde{R}^{-1}|
= |\widetilde{R}^{-T} A ^T A 
\widetilde{R}^{-1} - I|.
\end{array}
$$
Going back to the $R$ factor of the $QR$ factorization we then have
the following corollary to Theorem~\ref{theo-errorsun}
\begin{theorem}\label{theo-error}
For $A \in  \RR ^{n \times n}$ an invertible matrix, let $R$ 
be the $QR$ factor of $A$. Let $\widetilde{R}\in  \RR ^{n \times n}$
be upper triangular and invertible, and 
\begin{equation} \label{neq-defG}
G = |\widetilde{R}^{-T} A ^T A 
\widetilde{R}^{-1} - I|.
\end{equation}
Then if 
\begin{equation} \label{neq-testrho}
\rho (G) < 1,
\end{equation}
we have 
\begin{equation} \label{neq-errorH}
|\widetilde{R}-R| \leq \text{\rm triu} (G (I-G)^{-1}) |\widetilde{R}|.
\end{equation}
\end{theorem}
\begin{proof}
Since $\widetilde{R}$ is invertible,
$\widetilde{B}=\widetilde{R}^T\widetilde{R}$ is positive
definite, the same holds for $B=A ^T A$. By construction  
$R$ and $\widetilde{R}$ are the Cholesky factors of 
$B$ and $\widetilde{B}$. It suffices to apply
Theorem~\ref{theo-errorsun} for concluding.
\end{proof}

Few things are known about the (mathematical) quality of   
Bound~(\ref{neq-errorH}) over~$\RR$.
Furthermore, both additional method and arithmetic errors will be introduced for the finite
precision evaluation of the bound. Additional method errors will be 
introduced especially for calculating certified bounds for
$\widetilde{R}^{-1}$ and 
$\text{\rm triu}(G (I-G)^{-1})$
(see Section~\ref{sec-tow}). Additional arithmetic errors 
will be introduced by the finite precision itself.
All together we produce an error bounding algorithm that is not
fully 
analyzed, the experiments of
Section~\ref{subsec-comput}
will however give a precise idea of its practical behaviour and effectiveness. 
For illustrating Bound~(\ref{neq-errorH}) over~$\RR$,
let us consider some examples that show that  
Theorem~\ref{theo-error} leads to accurate bounds. The calculations
have been done in Maple~\cite{Maple10}, either exactly or with high precision,
then rounded for the presentation. 
Let $H = \text{\rm triu} (G (I-G)^{-1})$ such
that~(\ref{neq-errorH}) is  
$|\widetilde{R}-R| \leq H |\widetilde{R}|$.

On the matrices used for 
Figure~3.1  ({\tt randsvd}, $n=200$),
with $\widetilde{R}$ computed using 64~bits floating point
numbers via the Modified Gram-Schmidt
algorithm,   we typically get the following. 
For $A$ with $\text{cond}(R^{-1}) \approx 10^5$, 
the infinity norm of the error matrix is 
$\|H\|_{\infty} \approx 2 \times 10^{-9}$.
This leads to the knowledge that 
$\widetilde{R}$ approximates $R$ with (relative) accuracy $\approx
10^{-10}$.
The accuracy of $\widetilde{R}$ is about  
$10^{-13}$ for the diagonal entries, and the 
diagonal error estimation is only in a factor of $2$ from the
true diagonal error.    
If $\text{cond}(R^{-1}) \approx 4 \times 10^{13}$
then $\|H\|_{\infty} \approx 3 \times 10^{-3}$, and  $R$ is known 
with accuracy about $10^{-2}$ ($2 \times 10^{-5}$ on the diagonal). 
The ratio between the estimation and the true error is less than $4$
on the diagonal. 
Again, we will certainly loose accuracy with  our finite precision 
implementation, but keep a very satisfying overall
behaviour.  
Consider also the matrix quoted from~\cite[Eq.\,5.4]{ChPa01}:
$$
A_1 = \left[ \begin {array}{cc}  1& 1-10^{-10} 
\\\noalign{\medskip} 1& 1 + 10^{-10}\end {array} \right],
$$
with $\text{cond}(R^{-1}) \approx 2 \times 10^{10}$.  We compute the matrix 
$\widetilde{R}$ in Matlab~\cite{matlab06}, and obtain over $\RR$ the
error bound:
\begin{equation}\label{theoA1}
|\widetilde{R}-R| \approx 
\left[ \begin {array}{cc} { 9.7\times 10^{-17}}&-{
 1.3\times 10^{-16}}\\\noalign{\medskip} 0&{ 3.7
\times 10^{-17}}\end {array} \right]
\leq 
 \left[ \begin {array}{cc} { 3.5\times 10^{-12}}&{ 3.5
\times 10^{-12}}\\\noalign{\medskip} 0&{ 7.4\times 10^{-17}}
\end {array} \right].
\end{equation}
The matrix $R$ is known with (relative) accuracy about $2.5 \times
10^{-12}$ on the first row, and $5.25 \times
10^{-7}$ for $r_{22}$. On the first row the error is overestimated
by a factor about $3.6 \times 10^4$. Notwithstanding
the fact that the accuracy of
the bound produced by Theorem~\ref{theo-errorsun}
 is penalized by the particular form of the matrix,
the estimation of the accuracy of $\widetilde{R}$ remains very good. 
Now let $A$ be the random $3 \times 3$ integer matrix 
$$
A_2 = \left[ \begin {array}{ccc} - 60& 28& 51\\\noalign{\medskip}-
 24&- 35&- 89\\\noalign{\medskip} 37& 51&- 23\end {array}
 \right].
$$ 
We look at Bound~(\ref{neq-errorH}) when perturbing only the second
row of the exact $R$ and get: 
\begin{equation}\label{theoA2}
|\widetilde{R}-R| = 
\left|\left[ \begin {array}{ccc}  0 & 0& 0\\\noalign{\medskip}0&
 0.0071& -0.0052\\\noalign{\medskip}0&0& 0
\end {array} \right]\right| \leq  \left[ \begin {array}{ccc}  0 &
 0 & 0
\\\noalign{\medskip} 0& 0.014204 & 0.023087 
\\\noalign{\medskip} 0& 0& 2.9 \times 10 ^{-6}\end {array} \right].
\end{equation}
The estimator computes the errors very well on the first and the
second row. We think that the dummy error estimated for $r_{33}$ 
is a repercussion of the perturbation of row two. 
In next section we review the different quantities that are involved 
in Theorem~\ref{theo-error} with the aim of looking at first 
implementation aspects. 


\section{Toward an implementation}   \label{sec-tow}

Theorem~\ref{theo-error} is the foundation of our error bounding
algorithm. It involves several quantities that need further
study before deriving an implementation in Section~\ref{sec-errorR}. 
We decompose the computation of  
the bound on $|\widetilde{R}-R|$ into four principal tasks. 
We need to: 1) check  that $\widetilde{R}$ is invertible; 2) compute
a bound on $G$; 3) check that $\rho (G) < 1$; and 4) 
bound $H=\text{\rm triu}(G (I-G)^{-1})$. 
We recall that at this point, only $A$ and $\widetilde{R}$ are
known. 

\subsection{Invertibility check of $\widetilde{R}$} \label{subsec-checkR}
For dealing with $\widetilde{R} ^{-1}$ in a certified way, which is clearly  a
non trivial question in finite precision, we use the verification solution of Oishi and
Rump~\cite{OiRu02}. 
We compute 
a purely numerical approximate inverse $V \approx  \widetilde{R}
^{-1}$ (by numerical triangular inversion). 
Then we know from~\cite{OiRu02} that, if 
\begin{equation} \label{eq-testR}
\|\widetilde{R}V-I\|_{\infty} < 1,
\end{equation}
then $\widetilde{R}$ is invertible.

\subsection{Bounding $G$}\label{subsec-boundG}

For bounding $G$, and dealing with the unknown inverse of
$\widetilde{R}$, 
we are also inspired by~\cite{OiRu02}, and introduce 
$W=\widetilde{R}V$ ($\approx I$). We have  
$$
\begin{array}{ll}
G & = |\widetilde{R}^{-T} A ^T A 
\widetilde{R}^{-1} - I| \\
&=  |(W ^{-T} W ^T) \widetilde{R}^{-T} A ^T A 
\widetilde{R}^{-1} (W W ^{-1}) -  (W ^{-T} W ^T)(W W ^{-1})|\\
&= |W ^{-T}( V^T A ^T A V -   W ^TW) W ^{-1})|
\leq |W ^{-T}|\cdot|V^T A ^T A V -   W ^TW|\cdot| W ^{-1}|.
\end{array} 
$$
In the inequality above, if $\widetilde{R}$ is close to $R$ and $V$
is close to $\widetilde{R}^{-1}$, then both $V^T A ^T A V$ and $W
^TW$ are close to identity. Hence it is natural to pursue with:
$$
\begin{array}{ll}
G & \leq |W ^{-T}|\cdot|V^T A ^T A V - I + I -   W ^TW|\cdot| W ^{-1}|\\
&\leq |W ^{-T}|\cdot|(V^T A ^T A V -I) - (W ^TW-I) |\cdot| W
^{-1}|
\end{array} 
$$
which gives
\begin{equation} \label{eq-computG}
G \leq |W ^{-T}|\cdot(|(V^T A ^T A V -I)|+|(W ^TW-I)|)\cdot| W
^{-1}|.
\end{equation}
We will use~(\ref{eq-computG}) for computing a certified bound for
$G$. 
The products involving $A$, $\widetilde{R}$, $V$, 
and $W =  \widetilde{R}V$ will be bounded directly by interval
techniques. It remains to bound 
$| W
^{-1}|$. We expect $W$ to be close to $I$, and may use a
specific approximation.  
We have $|W^{-1}|= | (I-(I-W))^{-1}|$ (see~\cite[Intro.]{OiRu02}).
Then, when $\widetilde{R}$ is invertible,
$$
\begin{array}{ll}
|W^{-1}| & = | I + (I-W) + (I-W)^2 + \ldots |\\
& = | 2 I - W + (I-W)^2( I + (I-W) + (I-W)^2 + \ldots )|\\
& \leq | 2 I - W| + |(I-W)^2|\cdot |I + (I-W) + (I-W)^2 + \ldots|\\
& \leq | 2 I - W| + {\mathcal M}(\|I-W\|^2_{\infty}/(1-\|I-W\|_{\infty}))
\end{array} 
$$
where ${\mathcal M}(x)$ for $x \in \RR$ denotes the matrix whose all
entries are equal to $x$.
Here we have used the fact that the entries
of $|I-W|^2\cdot|I + (I-W) + (I-W)^2 + \ldots|$ are bounded by the
infinity norm. Since $W$ is triangular,  
it follows that 
\begin{equation} \label{eq-computabsinv}
|W^{-1}| \leq | 2 I - W| + \frac{\|I-W\|^2}{1-\|I-W\|_{\infty}}\cdot
\text{\rm triu}(\text{\rm 1}_n \cdot\text{\rm 1}^T_n)
\end{equation}
where $\text{\rm 1}_n$ is the column vector with all entries equal
to $1$. Note that the invertibility check~(\ref{eq-testR})
ensures that  $1-\|I-W\|_{\infty} >0$.
The absolute value $|W^{-1}|$ could have been bounded directly using
$1/(1-\|I-W\|_{\infty})$, but introducing the infinity norm only in
the second order terms leads to a much better bound 
in our experiments. 

The matrix manipulations we have done for obtaining~(\ref{eq-computG})
and~(\ref{eq-computabsinv}) follow some keys to the design
of verification methods. We especially refer
to~\cite[p.\,211]{Rum05} where the introduction of small factors is recommended.   
We have introduced  
the matrices $V^T A ^T A V-I$ and $W
^TW-I$ 
whose absolute bounds are expected to be small 
when  $\widetilde{R} \approx R$ and $W \approx I$. 
On the other hand, in~(\ref{eq-computabsinv}), $|2I-W|$ is expected
to be close to $I$, and remaining terms are second order terms (see also the analysis for 
$\alpha$ in~\cite[\S5]{OiRu02}).

\subsection{Bounding the spectral radius of $G$}\label{subsec-boundrhoG}

For any consistent matrix norm we have $\rho(A) \leq \|A\|$. With
the above 
bound on $G$, 
we will simply test whether  
\begin{equation} \label{eq-computrho}
\|G\|_{\infty} < 1  
\end{equation}
for asserting that $\rho(G) <1$ in Theorem~\ref{theo-error}. This
test corresponds to the Gershg\"orin disks. It could certainly 
be sharpened in future versions of the certificate, 
see for instance the Cassini ovals in~\cite{BrMe94},
or the iterative estimation in~\cite{Rum06}.

\subsection{Bounding $|\widetilde{R}-R|$}\label{subsec-boundH}

Once a bound on $G$ is known it remains to bound $H=\text{\rm
  triu}(G (I-G)^{-1})$.
We have 
$$
G(I-G)^{-1}  =  G + G^2 + G^3 + \ldots = G + G^2 (I + G + G^2 + \ldots) \\
$$
and 
\begin{equation} \label{eq-computH}
\text{\rm
  triu}(G (I-G)^{-1}) 
\leq \text{\rm
  triu}(G) +  \text{\rm
  triu}\left(\frac{\|G\|_{\infty}^2}{1-\|G\|_{\infty}}\cdot 
\text{\rm 1}_n \cdot\text{\rm 1}^T_n \right).
\end{equation}
Since $G$ is expected to be small, $H = \text{\rm
  triu}(G (I-G)^{-1})$ is expected to be close to   $\text{\rm
  triu}(G)$.
Note that using the spectral radius check~(\ref{eq-computrho})
ensures that ${1-\|G\|_{\infty}}>0$.


\section{Error bounding algorithm for the $QR$ factor $R$} \label{sec-errorR}

Let $\FF$ be a set of floating point numbers such that the arithmetic
operations in~$\FF$ satisfy the IEEE~754 standard. 
$A$ and $\widetilde{R}$ are now matrices in
$\FF ^{n \times n}$. Since (finite) floating point numbers 
are rational numbers, $A$ and $\widetilde{R}$ can be seen as rational
matrices. Let $R \in \RR ^{n \times n}$ be the unknown $QR$ factor of $A$
(in general, the entries of $R$ are not in~$\FF$).
We carry the approach of  Section~\ref{sec-tow}
over to the floating point case for computing 
a floating point matrix $H$ such that $|\widetilde{R}-R| \leq H
|\widetilde{R}|$. 
The error matrix $H$ provided by Theorem~\ref{theo-error}
can be computed modulo the two checks~(\ref{eq-testR})
and~(\ref{eq-computrho}), and using the inequalities~(\ref{eq-computG}), (\ref{eq-computabsinv}),
and~(\ref{eq-computH}).
These checks and inequalities only involve matrix multiplications,
additions, subtractions, and divisions by a scalar. 
After explaining the basic techniques we use for computing certified
bounds in floating point arithmetic, we present the error bounding
algorithm and demonstrate its effectiveness on various examples.

\subsection{Certified bounds for floating point matrix expressions} \label{subsec:certiftech}

We denote by $\text{\rm fl}(x)$ the value of an arithmetic
expression $x$ computed by floating point arithmetic in $\FF$.
For instance, for $a,b \in \FF$, $\text{\rm fl}(a+b\times c)$ denotes the 
result in $\FF$ with the addition and the mutiplication performed in
floating point arithmetic.  In the text, an arithmetic
expression on floating point numbers denotes the exact value in
$\RR$. For instance $a+b \in \RR$ is the result of the addition in $\RR$.
The abolute value, the max, and the negation are exact operations: for $a,b \in
\FF$, $\text{\rm
  fl}(|a|)= |a|$, $\text{\rm
  fl}(\max\{a,b\}) = \max\{a,b\}$, $\text{\rm
  fl}(-a)= -a$.

Thanks to the IEEE~754 standard, we can use the
possibility of changing the rounding mode for computing certified
bounds. We essentially follow Rump's approach for implementing 
verified matrix operations~\cite{Rum05}, and Oishi and Rump~\cite{OiRu02}.
We use the statements ``setround(down)'' and
``setround(up)''\footnote{
{\tt  fesetround(FE\_DOWNWARD)} and {\tt fesetround(FE\_UPWARD)}
in C language.}.
 All operations after a statement ``setround(down)'' or ``setround(up)''  are rounded
 downwards or upwards, respectively, until the 
next call to setround. For two floating point numbers $a$ and $b$, a
bound $r$ on 
$|a~op~b|$ for $op \in \{+,-,\times,\div\}$
may be computed as follows. The program 
\begin{equation}\label{eq-roundmax}
\begin{array}{ll}
\text{setround(down)};
&\underline{r}=\text{\rm fl}(a~op~b)\\
\text{setround(up)};
&\overline{r}=\text{\rm fl}(a~op~b); ~~r = \max \{|\underline{r}|,|\overline{r}|\}
\end{array}
\end{equation}
leads to $\underline{r}$ and $\overline{r}$ such that 
$
\underline{r} \leq a~op~b \leq \overline{r},
$
and to $r \in \FF$ such that 
$
|a~op~b| \leq r, 
$ 
for any $a$ and $b$, and any $op$. 
The IEEE standard ensures that $\underline{r}$ and $\overline{r}$
are the best possible bounds in $\FF$.
This may be extended to the matrix operation 
$A \times B - C$ with  $A,B,C \in \FF ^{n \times n}$. If $A \times B$  
is implemented using only additions and multiplications, then the
program 
\begin{equation}\label{eq-ABC}
\begin{array}{ll}
\text{setround(down)}; 
&\underline{R}=\text{\rm fl}(A\times B-C)\\
\text{setround(up)}; 
&\overline{R}=\text{\rm fl}(A \times B -C); ~~R = \max \{|\underline{R}|,|\overline{R}|\}
\end{array}
\end{equation}
where the maximum is taken componentwise, provides $\underline{R}
\leq A \times B -C \leq \overline{R}$, and $R \in \FF ^{n
  \times n}$ such that 
$|A\times B-C| \leq R$.
For bounding more general matrix expressions we will use a
midpoint-radius matrix representation (we refer
to~\cite[\S10.9]{Rum05}).
Assume that $M$ and $N$ are two matrices known 
to be in intervals $[\underline{M},\overline{M}]$
and $[\underline{N},\overline{N}]$, respectively.
The intervals are for instance obtained by a computation of
type~(\ref{eq-ABC}).
Then the program~\cite[Fig.\,10.22]{Rum05}:
\begin{equation}\label{eq-MN}
\begin{array}{ll}
\text{setround(up)};&  \text{\rm m}_M = \text{\rm fl} ((\overline{M}
-\underline{M})/2);~~\text{\rm r}_M = \text{\rm fl} (\text{\rm m}_M -
\underline{M})\\
&\text{\rm m}_N = \text{\rm fl} ((\overline{N}
-\underline{N})/2);~~\text{\rm r}_N = \text{\rm fl} (\text{\rm m}_N -
\underline{N})\\
\text{setround(down)};& \underline{R} = \text{\rm fl} (\text{\rm m}_M \times
\text{\rm m}_N-I)\\
\text{setround(up)};& \overline{R} = \text{\rm fl} (\text{\rm m}_M \times
\text{\rm m}_N-I)\\
& R = \text{\rm fl} \left(\max \{|\underline{R}|,|\overline{R}|\} + |\text{\rm
  m}_M|\times \text{\rm r}_N + \text{\rm r}_M \times \left(
|\text{\rm
  m}_N| + \text{\rm r}_N
 \right)\right)
\end{array}
\end{equation}
computes $R$ such that $|M\times N-I| \leq R$.
Both~(\ref{eq-ABC}) and~(\ref{eq-MN}) allow to use fast matrix
routines such as the BLAS ones (see the general
discussion in~\cite[\S10.9]{Rum05})
The number of operations in $\FF$ needed is  
$2$ and $4$ matrix products, respectively.

Other matrix operations that we will perform  are additions,
products, and divisions by scalars for matrices with positive
entries (absolute values essentially). We   
also compute infinity norms. With no subtraction involved, certified
bounds can be computed using
directed rounding. From~(\ref{eq-roundmax}), upper bounds for
these computations are obtained 
by evaluating the floating point expressions 
after a ``setround(up)'' statement. 
For upper bounds on divisions by a floating point number $1-g$, we first compute  
upper bounds for $-(g-1)$ and $1/(g-1)$. 

Other approaches for certified matrix computations could be
considered. We refer to Rump~\cite{Rum05} for a general discussion
on this topic, and for the efficiency of the approach chosen here.

\subsection{Computing an error bound}

For $A$ and $\widetilde{R}$ in
$\FF ^{n \times n}$, $\widetilde{R}$ upper triangular, 
we follow Section~\ref{sec-tow} for computing 
a floating point matrix $H$ such that $|\widetilde{R}-R| \leq H
|\widetilde{R}|$. All operations are done in the given floating
point number set $\FF$. For simplifying the presentation we often forget
the costs in $O(n^2)$.

The first step is the 
computation of $V \approx \widetilde{R}^{-1}$. Such a triangular
matrix inversion is done in $n^3/3$
operations~\cite[Ch.\,14]{Hig02}. We then compute $\underline{W}$ and
$\overline{W}$ for $W=\widetilde{R}V$ by two triangular matrix
products, this is done in $2n^3/3$ operations. 
This dominates the cost for checking that $\widetilde{R}$ is
invertible by bounding $|W-I|$
using~(\ref{eq-ABC}), and by the infinity norm test~(\ref{eq-testR}). 
Re-using $\underline{W}$ and
$\overline{W}$,
a bound on $|W ^{-1}|$ is then computed
using~(\ref{eq-computabsinv}) in $O(n^2)$ operations.
The latter uses~(\ref{eq-ABC}) 
for $|W-2I|$, and computes a bound with positive matrices using
directed upwards
rounding.
We now look at bounding $G$ using~(\ref{eq-computG}).
Since $G$ is symmetric we restrict ourselves to counting the
operations for calculating 
 the upper triangular part. 
With $W \in [\underline{W},\overline{W}]$  one can
bound $|W^TW-I|$ using~(\ref{eq-MN}) in four matrix
products. Since $W$ is upper triangular, and $W^T$ is lower triangular,
the bound is obtained in $4n^3/3$
operations. 
We then use~(\ref{eq-ABC})
and~(\ref{eq-MN}) for computing an interval for $AV$ in $2n^3$
operations (two dense $\times$ triangular matrix products), and for 
bounding  $|V^TA ^T A V^T-I|$ in $4n^3$ operations (four dense
products resulting in a symmetric matrix). A bound on $G$ is deduced by
operations on matrices with positive entries in $4n^3/3$ operations. 
The latter is  essentially two   
dense $\times$ triangular matrix products with a symmetric result. 
Once a bound on $G$ is known, testing its spectral radius
by~(\ref{eq-computrho}) costs $O(n^2)$ operations. 
$G$ has positive entries,  a bound on the
error matrix $H$ can then be computed by directed towards rounding
using~(\ref{eq-computH}) also in  $O(n^2)$ operations. 

We summarize this analysis, and take into account the final matrix product $H
|\widetilde{R}|$ in the following result. 

\begin{theorem} \label{theo-maincost} 
Let $A \in \FF ^{n \times n}$, and $\widetilde{R} \in \FF ^{n \times
  n}$ upper triangular be given. 
The error bounding algorithm computes a matrix 
$F\in \FF ^{n \times
  n}$ such that $|\widetilde{R}-R| \leq F$, where $R$ is the unknown $QR$ 
factor of $A$, in $10n^3+O(n^2)$ floating point
operations. 
\end{theorem}

A $QR$ factorization typically costs $2n^3 +O(n^2)$
(Gram-Schmidt or Householder approaches) or $3n^3 +O(n^2)$ (using
Givens rotations). Hence we are able to compute a certified 
error bound $|\widetilde{R}-R|$ at the cost of only five
approximate factorizations. 
We have implemented the algorithm in C language. 
The error bounding program 
takes in input  two floating point matrices $A$ and
$\widetilde{R}$
and always returns a matrix $F$.
The entries of $F$ are finite (positive) floating 
numbers if the program is able to certify that $\widetilde{R}$
is invertible, that the spectral radius of $G$ is less than
one, and if no overflow is produced. 
Otherwise, the entries of $F$ may be equal to infinity. 


\subsection{Computational results}\label{subsec-comput}

The results we present here correspond
to the application  of Theorem\ref{theo-maincost} with $64$~bits
floating point numbers. 
In this section and in~Section~\ref{sec-reduced} the condition
numbers and the ``true errors''
have been computed with high precision 
using Mpfr~\cite{mpfr06}.
For several types of matrices, 
we study the behaviour of the certified error bound by looking at
its value and its accuracy (with respect to the
true error), especially when the dimension and the condition number
increase. 
We mainly focus on the exponent $k$ such that relative error is in 
$10^{-k}$, $k$ expresses the number of significant decimal
digits we certify for the entries of $\widetilde{R}$.  
Let us first come back on the examples of Section~\ref{sec-sun}. 
On the matrix $A_1$, and $\widetilde{R}$ from Matlab, we compute the bound 
$$
|\widetilde{R}-R| \leq 
 \left[ \begin {array}{cc} { 6.7\times 10^{-11}}&{ 6.7
\times 10^{-11}}\\\noalign{\medskip} 0&{ 5\times 10^{-16}}
\end {array} \right].
$$
Comparing to~(\ref{theoA1}), we see that the finite precision
estimator we propose is only slightly overestimating the  
best bound that could be obtained by the method. 
On the matrix $A_2$, and the corresponding perturbation of the exact
$R$ we get:
$$
|\widetilde{R}-R| \leq  
\left[ \begin {array}{ccc}  8.8 \times 10^{-6} &  9.52 \times 10^{-6}& 1.96 \times 10^{-6}\\
\noalign{\medskip} 0& 0.014207 & 0.023098 
\\\noalign{\medskip} 0& 0& 1.16 \times 10 ^{-5}\end {array} \right].
$$
The ``large'' perturbation of the second row is detected very
accurately. 
For next results, $\widetilde{R}$ is computed
with the Modified Gram-Schmidt algorithm using $64$~bits numbers 
as for the estimator. Our tests use ten matrix samples.

\begin{center}
\includegraphics[scale=0.5]{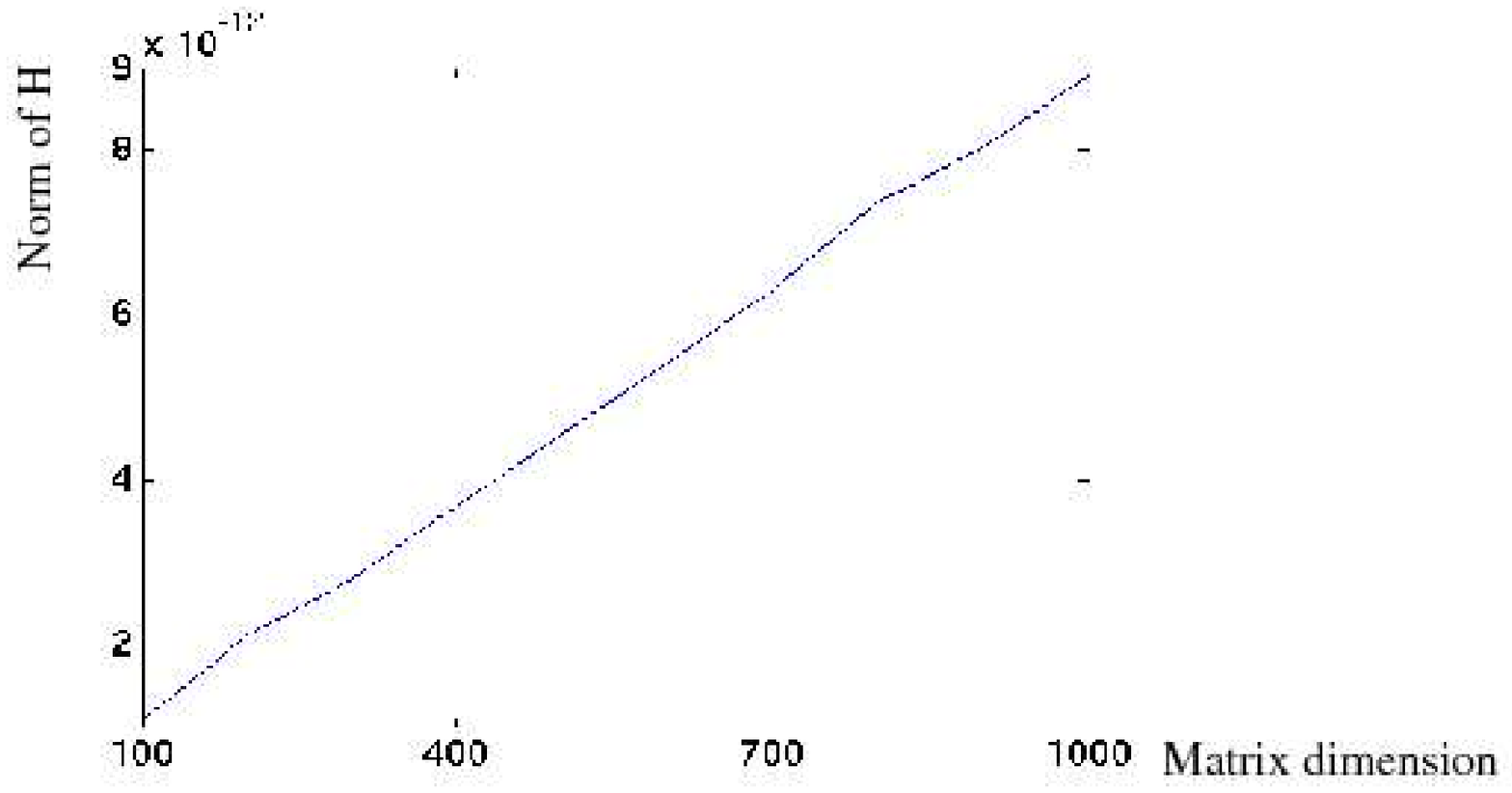} 
{\footnotesize
~\\
\begin{minipage}{13.4cm}\begin{center}
{\footnotesize Figure 6.1: Certified $\|H\|_{\infty}$  
for random matrices $A$ with  
$\kappa _{2}(A) \approx 10^3$.} \end{center}
\end{minipage}
}
\end{center}
We first illustrate the {\em value of the certified bound with respect to
the dimension}.  
Figures~6.1 and~6.2 are for random input matrices $A$
(of {\tt randsvd} type~\cite[Ch.\,28]{Hig02}). 
\begin{center}
\includegraphics[scale=0.5]{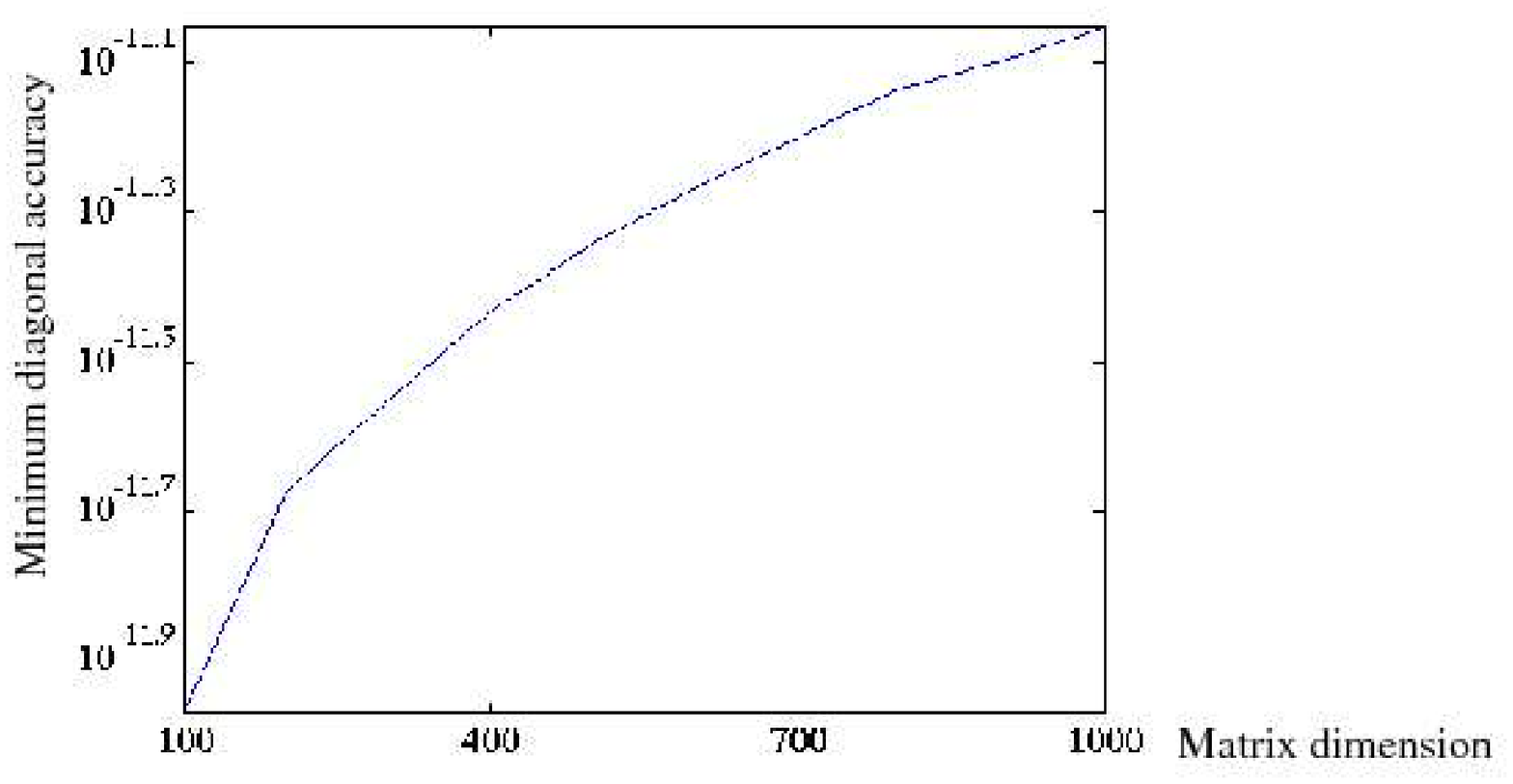} 
{\footnotesize
~\\
\begin{minipage}{13.4cm}\begin{center}
{\footnotesize Figure 6.2: Certified maximum relative error on ${R}$
  for  random matrices $A$ \\such that 
$\kappa _{2}(A) \approx 10^3$ ($y$ axe with logarithmic scale).} \end{center}
\end{minipage}
}
\end{center}
We keep the condition
number almost constant when the dimension increase. We draw the
infinity norm of $H$ such that $|\widetilde{R}-R| \leq H
|\widetilde{R}|$, and the certified maximum relative error on the
diagonal of $\widetilde{R}$, we mean $\max _i
|\widetilde{r}_{ii}-r_{ii}|/|\widetilde{r}_{ii}|$. We see that
$\|H\|_{\infty}$ increases linearly with $n$. The loss of accuracy
on the diagonal is approximately quadratic in $n$ (we use a logarithmic scale
for the $y$ axe on Figure~6.2). 
Such small increase rates---that are typical of numerical
algorithm forward errors themselves---demonstrate a first aspect of
the effectiveness of our finite precision bounds.   
The certified general maximum error $\max _{ij}
|\widetilde{r}_{ij}-r_{ij}|/|\widetilde{r}_{ij}|$ increases
faster. It typically grows from $10 ^{-7}$ to $10^{-5}$ for
the dimensions considered here. 
We need further investigation for a better understanding of the latter
behaviour, especially of the influence of  
the product $H|\widetilde{R}|$, 
and of the magnitudes in $R$. 
Note also that for the two latter figures, $\text{\rm
  cond}(R^{-1})$ is  sligthly growing, and the growth of the
estimation
depends on the true error itself.  

We discuss next the {\em accuracy of the 
certified bound with respect to the exact error} (not the quality of
the $QR$ algorithm itself). 
In addition to above {\tt randsvd} matrices we also consider 
random integer matrices with entries of absolute values less than $1000$.
On these two types of matrices we obtain similar results. 
The condition numbers  $\kappa _{\infty}(A)$ 
are varying from about $10^4$ to $10^6$.  
On random integer matrices of dimension $1500$, the maximum exact
 relative error on
 $R$ 
has order $10^{-10}$
to $10^{-9}$. We are able to certify this error by returning an error bound of
order $10^{-6}$
to $10^{-5}$. 
{\em With respect to the dimension}, we observe 
that the fast certified bound  
overestimates the componentwise error by a factor 
of order about $10^3$ for $n=200$ to 
about $10^5$ for $n=1500$. 
Restricted to the diagonal entries, the overestimation goes from 
about $10^2$ to less than $10^4$. 
This shows that even with condition numbers and dimensions that
can be here quite large,
we are able to certify at least four or five significant decimal digits for
every entries of $R$, and at least $9$ digits on the diagonal
(where the error itself is much smaller in general). 
On matrices with small condition number (generated using Matlab {\tt
  gallery('orthog')}~\cite[Chapter\,28]{Hig02}) the quality of the
certified bound may be remarkably small and stable with respect to the
dimension. For dimensions between $60$ and $500$, and 
$\text{\rm cond}(R ^{-1})\approx 3$ ($\kappa _{\infty} \leq 200$), we most of the time obtain an overestimation
between $15$ and $22$ (and more than $12$ certified significant
decimal digits in $\widetilde{R}$).

We may now ask the question of the sensitivity of the {\em quality of the
certified error bound with
respect to the condition number of the input matrix}. 
We first report that 
the quality maybe be very good even for
matrices with  high condition number. 
For Figure~6.4
we use $A=QA_K \in \FF ^{n \times n}$. 
The matrices $Q$ are random orthogonal from the Matlab
{\tt gallery} function~\cite[Chapter\,28]{Hig02}. The matrices $A_K$ are 
Kahan upper triangular matrices with $a_{ii}= (\sin \theta)^{i-1}$, 
$a_{ij}= -(\sin \theta)^{i-1} \cos \theta$ for $j > i$, and $\theta=1.2$.
\begin{center}
\begin{minipage}{16cm}\begin{center}
{\footnotesize
$$
\begin{array}{|c|c|c|c|c|c|c|c|} \hline 
\text{\rm Dimension} & 10 & 20 & 30 & 40 & 50 & 60 & 70  \\  
\kappa _{\infty}(A) & 10^2 & 1.3 \times 10^4 & 1.1 \times 10^6 &
7.8 \times 10^7 & 4.8 \times 10^9 & 
2.8 \times 10^{11} & 1.5 \times 10^{13}\\\hline  
\text{\rm Bound/error} & 45 & 106 & 281 & 161 & 103 &140 & 152\\
\text{\rm Certified digits in~} \widetilde{R} & 14 & 12 & 10 & 9 & 7 & 5 & 4 
\\\hline 
\end{array}
$$
Figure 6.4: Ratio of the certified relative error bound and the true
error (max) \\on
Kahan matrices, and number of significant decimal digits certified
in $\widetilde{R}$. 
}\end{center}
\end{minipage}
\end{center}
However, in general, the quality of the bound may
depend on the condition number. Consider for instance the 
ratio of the certified relative error bound and the true error (max)
for
small matrices ($n=10$).  
For a Chebyshev Vandermonde-like (nearly orthogonal, $\kappa
_{\infty} \approx 13$), the ratio is about $11$. We have 
a ratio about $14$ for 
Toeplitz and symmetric positive definite matrices 
($\kappa
_{\infty} \approx 700$). 
On the Pascal matrix ($\kappa
_{\infty} \approx 8 \times 10^9$) we get a ratio about $25$, and
about $1600$ for the Hilbert matrix  ($\kappa
_{\infty} \approx 3.5 \times 10^{13}$).
Figure~6.5 is more general. The overestimation of 
certified error bound seem to increase quite slowly with the
condition number. 
\begin{center}
\includegraphics[scale=0.5]{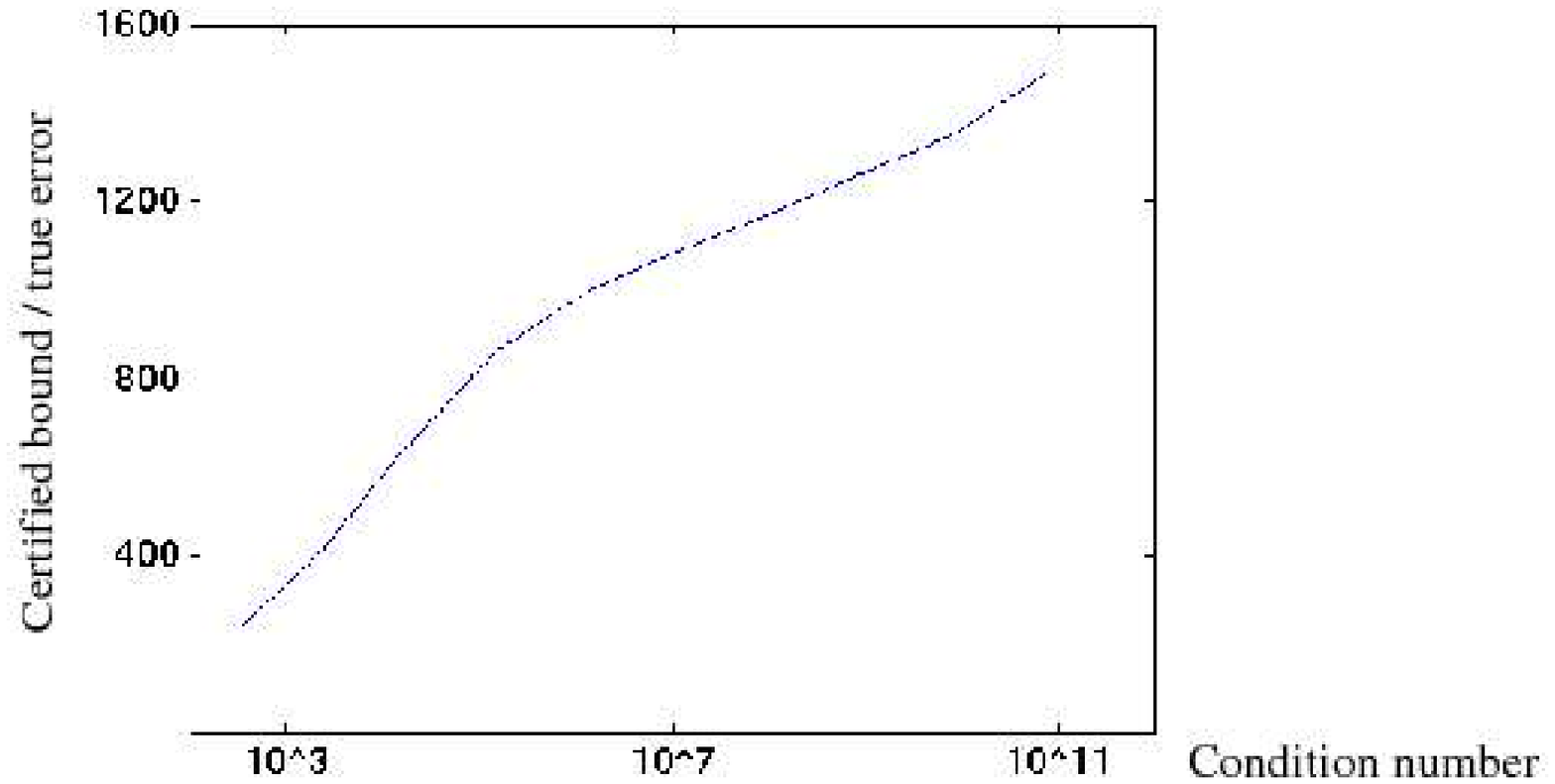} 
{\footnotesize
~\\
\begin{minipage}{13.4cm}\begin{center}
{\footnotesize Figure 6.5: Ratio of the certified relative error
  bound and the true error (max)\\
with respect to $\kappa_{\infty}(A)$ on
{\tt randsvd} matrices of dimension $n=200$.} \end{center}
\end{minipage}
}
\end{center}

We see that the limits of our algorithm, we mean the conditions in
which it is returning finite bounds, are clearly linked 
with the numerical properties of $A$. 
Let us give two examples for the impossibility 
to certify the spectral radius using~(\ref{eq-computrho}). 
We return finite bounds for the error on every entries of
$\widetilde{R}$ for the Pascal matrix of dimension $14$ ($\kappa
_{\infty} \approx 3.8 \times 10^{14}$, $\|G\|_{\infty}\approx
0.06$). For $n=15$ the algorithm produces infinity bounds. 
On random {\tt randsvd} matrices of dimension $40$, the algorithm
is effective until $\kappa
_{\infty} \approx 3 \times 10^{14}$ with $\|G\|_{\infty}\approx
0.9$. Note
that in double precision, with relative rounding unit $2^{-53}$ (the
backward error is larger in general),
and for a relative forward error less than $1$,  
 the rule of thumb~(\ref{thumb})
advocates for a condition number less than $10^{16}$.

The certified bound is computed with finite precision, hence inherently, 
it overestimates the true error. However, for realistic dimensions
and condition numbers (with respect to the precision), the
overestimation is mastered. It follows that in general, many significant digits are
certified in the approximate $QR$ factor $\widetilde{R}$.   
The latter is a key to the application of the
fast bound to the reducedness certificate. 


\section{A certificate for LLL reducedness} \label{sec-reduced}

To an $n\times n$ integer matrix $A$ we associate the Euclidean 
lattice ${\mathcal L}$ generated by the columns  
$(a_j)$ of $A$. About lattices the reader may refer for instance to
Cohen's book~\cite{Coh95}. 
Since the seminal 
Lenstra-Lenstra-Lov\'asz algorithm~\cite{LLL82}---whose range of
application is exceptional---the lattice basis
reduction problem receives much attention. 
In particular, floating-point variants 
that lead to very fast reduction approaches have been 
invented. See the work of Nguyen and Stehl\'e~\cite{NgSt05,Ste05},
of Schnorr~\cite{Sch06}, and references therein. 
Most of floating point variants 
lead to powerful heuristics, especially \`a la
Schnorr-Euchner~\cite{ScEu94}, that are implemented (often with
improvements) in most of
  computer algebra and number theory systems. 
Our aim here is not to study the basis reduction itself.
We focus on the reducedness. Indeed, 
a fast heuristic may not certify that the output basis 
is reduced (still working very well), and it is 
worthwhile to study the problem of checking {\em a posteriori} 
whether a given basis is reduced or not. 
The notion of reduction we consider is the 
LLL reduction~\cite{LLL82}.

We propose here an algorithm that takes in input an invertible  matrix $A \in \ZZ
^{n \times n}$, and tests the LLL reducedness of the basis formed
by the columns of $A$. 
In the Introduction we have seen that this consists in testing the 
two conditions~(\ref{eq:defproper}) and~(\ref{eq:deflovasz}).
Let $R$ be the $QR$ factor of $A$. 
If the $a_j$ are proper, we mean 
\begin{equation} \label{eq:proper}
|r_{i,j}|/r_{i,i}\leq \eta,~1
\leq i < j \leq n,
\end{equation}
and if the Lov\'asz conditions 
\begin{equation} \label{eq:lovasz}
\sqrt{\delta - \left(r_{i,i+1}/r_{i,i}\right)  ^2} ~r_{i,i} \leq r_{i+1,i+1},
~1\leq i \leq n-1,
\end{equation}
are satisfied, 
then the basis $a_1, a_2, \ldots, a_n$ of ${\mathcal L}$ is called LLL reduced with
parameters $(\delta, \eta)$.
The latter satisfy
$1/4 < \delta \leq 1$ and $1/2 \leq \eta < \sqrt{\delta}$.

The principle of the algorithm  
is to compute an approximate
$\widetilde{R}$ together with error bounds (using  the floating point
algorithm of Section~\ref{sec-errorR}), then to test~(\ref{eq:proper}) and~(\ref{eq:lovasz}). 

The entries of $A$ are integers of arbitrary size (our
implementation relies on Gmp~\cite{gmp06}). Therefore 
the entries of $A$ may not be represented exactly by elements in
$\FF$. Nevertheless, 
for the computation of an approximate 
$\widetilde{R}$ we may take $\widetilde{A}$
by direct conversion to $\FF$. Since the error is very small 
and $\widetilde{R}$ will be  an approximation anyway, 
this
does not really influence the quality of subsequent computations. Then
$\widetilde{R}$ is computed by the Modified Gram-Schmidt algorithm. 
Once $\widetilde{R}$ is known we apply Theorem~\ref{theo-maincost} 
for computing a certified error bound. 
The only expression that has to be bounded with $A$ involved
is in~(\ref{eq-computG}), where 
the computation of $AV$ using the program~(\ref{eq-ABC}) is needed.
The problem of conversion to $\FF$ is solved here by rounding upwards and
downwards 
during the conversion integer to
floating point. We mean that we introduce a small interval 
such that $A\in [A_{-}, A_+]$ with 
$A_-, A_+ \in \FF ^{n \times n}$
 (see the certified techniques in Section~\ref{subsec:certiftech}),
 and we evaluate 
 $A_-V$ and $A_+V$ in~(\ref{eq-ABC}). 
Therefore the error bound $F \in \FF ^{n \times n}$ 
we compute by Theorem~\ref{theo-maincost} is actually such that 
$|R-\widetilde{R}| \leq F$ for $R$ the $QR$ factor of any $A \in
[A_{-}, A_+]$. 

Once $F$ is known, for fixed $i$ and $j$, we test~(\ref{eq:proper})
by resorting to the bounding techniques  of Section~\ref{subsec:certiftech}:
\begin{equation}\label{eq-progprop}
\begin{array}{ll}
\text{setround(down)};
&\underline{\eta}=\text{\rm fl}(\eta); ~~t_i=\text{\rm
  fl}((r_{i,i}-f_{i,i})\times \underline{\eta} )\\
\text{setround(up)};
&t_j=\text{\rm fl}(|r_{i,j}|+f_{i,j})\\
& \text{\rm test} ~ t_j \leq t_i ?
\end{array}
\end{equation}
with temporary variables  $t_i$ and $t_j$.  Recall that the
diagonal entries of $R$ are positive. 
Similarly, for a fixed $i$, we test~(\ref{eq:lovasz}) using:
\begin{equation}\label{eq-proglov}
\begin{array}{ll}
\text{setround(up)}; & {t_i}=\text{\rm fl}(r_{i,i}+ f_{i,i});~ \overline{\delta}=\text{\rm fl}(\delta);\\
\text{setround(down)}; & t_{i+1}=\text{\rm fl}(r_{i+1,i+1}- f_{i+1,i+1}) \\
& t = \text{\rm fl}\left(((|r_{i,i+1}|- f_{i,i+1})/t_i)^2\right) -
\delta; ~t=-t;\\
\text{setround(up)};
&t=\text{\rm fl}(\sqrt{t}\times t_i)\\& \text{\rm test} ~ t \leq t_{i+1}?
\end{array}
\end{equation}
with temporary variables $t$ and $t_i$. 
In practice, for minimizing the cost induced by the changes of
rounding mode, loops are put between the setround instructions. 
In addition to the  $10n^3+O(n^2)$ operations for computing
$F$ using Theorem~\ref{theo-maincost}, the reducedness test 
essentially requires  $2n^3+O(n^2)$ operations
for computing an approximate factor $\widetilde{R}$. This gives the
following.
\begin{theorem} \label{theo-mainreduc} 
Let $A \in \ZZ ^{n \times n}$ invertible and parameters 
$(\delta, \eta)$ be given.
The reducedness certificate certifies  
in $12n^3+O(n^2)$ floating point
operations that the column lattice of $A$ is LLL reduced with parameters 
 $(\delta, \eta)$, or returns  ``failed''.
\end{theorem}
The reducedness is certified when the  error bound computed for 
$|\widetilde{R}-R|$ is finite, when no overflow or underflow occur
during the test, and when the basis is reduced.
The cost of the certificate is roughly the
one of six floating point
$QR$ 
factorizations.  Therefore in general, the reducedness test should be much
faster than the reduction process itself,  and may 
be appended  to any reduction heuristic program. 

Let us now report some experiments. As previously in the paper all
codes are run using $64$~bits floating point numbers. The effectiveness of the
certificate essentially relies on 
the effectiveness of the error bounding algorithm. 
We have manipulated lattices using Magma~\cite{magma06}, 
the LLL reduction implementation is based on the work of Nguyen and
Stehl\'e~\cite{NgSt05,Ste05}. 
The first family of reduced bases---matrices $A$---we consider are 
obtained by the reduction of $n \times n$ random integer matrices. 
The bases are reduced for the classical LLL parameters
$(\delta,\eta)=(3/4,1/2)$ in Figure~7.1, and $(\delta,\eta)=(0.99,0.5001)$
for a stronger reduction in Figure~7.2.
\begin{center}
\begin{minipage}{13.4cm}\begin{center}
{\footnotesize
$$
\begin{array}{|c|c|c|c|c|c|c|c|} \hline 
\text{\rm Dimension} & 40 & 200 & 500 & 1000   \\  \hline 
\kappa _{\infty}(A) & 4.7 \times 10^2 & 2.4 \times 10^4 & 1.8 \times 10^5 &
9 \times 10^5 \\ 
t_k-t=\min_i \{t_{i+1}-t\} \text{\rm ~in~(\ref{eq-proglov})}& 18 & 10 & 13 & 23\\
\text{\rm Certified~absolute~error~on~} \|a_k ^*\|_2 & 7.5 \times 10^{-12}
&  3 \times 10^{-10} & 1.5 \times 10^{-9}  & 1.2 \times 10^{-8}
\\\hline 
\text{\rm Certified~} \max_{ij} \mu _{ij} & 0.4997 & 0.499994 & 0.49991 & 0.49999\\
\text{\rm Max.~certified~relative~error~on~} |r_{ij}|
& 2.8 \times 10^{-11}& 8.6 \times 10^{-9}& 1.5 \times 10^{-7}& 3 \times 10^{-5}\\\hline
\end{array}
$$
Figure 7.1: Reducedness certificate output on $(3/4,1/2)$-reduced
bases from random integer matrices with entries on $10^3$ bits, 
$\max |a_{ij}| \leq 1000$.
}\end{center}
\end{minipage}
\end{center}
Since the numerical quality of the tested
bases is good ($\kappa _{\infty}(A) \leq 10^6$), the reducedness certificate 
is highly efficient. We mean that the
certified error is very small, and hence  the tests are passed
except in exceptional cases. 
Figures~7.1 and~7.2 for instance look at the smallest difference 
$t_k-t$ whose positiveness has to be certified
in~(\ref{eq-proglov}). The certificate has lots of room since the 
absolute errors on $t$ and $t_k = \|a_k^*\|_2$ are much smaller.  
Exceptional cases will rather occur when testing properness.
Indeed, testing
reducedness may be an ill-posed problem because of the possible equalities 
in~(\ref{eq:proper}) and~(\ref{eq:lovasz}). 
An ill-posed case with say $\eta=1/2$,
is for example a reduced basis with  $\mu _{ij}=1/2$ for some $i,j$.  
Therefore the algorithm will rather be used for certifying that 
a $(\delta,\eta)$-reduced basis is a $(\delta -\epsilon_1,\eta +\epsilon_2)$-reduced basis
for small $\epsilon_1, \epsilon_2$. The latter does really affect  
the relevant certified informations provided by the reduction. 
\begin{center}
\begin{minipage}{14cm}\begin{center}
{\footnotesize
$$
\begin{array}{|c|c|c|c|c|c|c|c|} \hline 
\text{\rm Dimension} & 40 & 200 & 500 & 1000   \\  \hline 
t_k-t=\min_i \{t_{i+1}-t\} \text{\rm ~in~(\ref{eq-proglov})}&  4.8 \times 10^{-2}& 7.7 \times 10^{-2}& 5.3 \times 10^{-2}&
7.3 \times 10^{-2}\\
\text{\rm Certified~absolute~error~on~} \|a_k ^*\|_2 & 9.4 \times 10^{-14}
&  6 \times 10^{-12} & 4 \times 10^{-11}  & 2 \times 10^{-10}
\\\hline 
\end{array}
$$
Figure 7.2: Reducedness certificate output on $(0.99,0.501)$-reduced
bases from random integer matrices with entries on $10$ bits, 
$\max |a_{ij}| \leq 10$.
}\end{center}
\end{minipage}
\end{center}

A second type of reduced bases on which we have run the certificate
comes from the problem of computing a good floating point coefficient
polynomial approximation to a function~\cite{BrCh07}. 
We have considered reduced bases with parameters  $(3/4,1/2)$.
These bases may have integer entries as large as $10^{80}$.
The certificate has always succeeded. 
On a $18 \times 18$ example, with $\kappa _{\infty}(A) \approx 4
\times 10^{12}$, the smallest difference $t-t_k$ has been around 
$2.4 \times 10^{76}$
with certified absolute error $1.95 \times 10^{62}$. 
The maximum of the $\mu _{ij}$ has been certified to be less than
$0.493$.
On an example with $n=31$ and $\kappa _{\infty}(A) \approx 8
\times 10^{13}$, 
we have  certified an absolute error $3.2 \times 10^{53}$ for 
$t-t_k \approx 1.7 \times 10^{67}$. 
On the latter example we have also checked that $\max \mu _{ij} \leq 0.4991$,
thanks to a  maximum relative error $|\widetilde{R}-R|$ certified to
be less than $0.2$ (only
$6 \times 10^{-15}$ on the diagonal).

The first main source of failure of the 
certificate is the failure of the error bounding algorithm when 
the precision is too small compared to the numerical quality of the 
tested basis. We have run the certificate on a third class of
reduced bases. These bases are obtained by the reduction of ``random'' (knapsack type)
 lattice bases in the  
sense of~\cite[\S3.4]{NgSt06}. In the experiments 
reported here, the non reduced bases have random integers 
of $10^3$ bits in the knapsack weight row. 
The reduced bases in input of the certificate (matrices $A$) are dense with
integers as large as $10^{45}$ for $n=75$, and  $10^{20}$
for $n=300$.
We use the parameters 
$(\delta,\eta)=(3/4,1/2)$ and $(\delta,\eta)=(0.99,0.5001)$.
The choice $(\delta,\eta)=(0.99,0.5001)$ 
produces better reduced bases as shown by $\kappa _{\infty}$
in 
Figure~7.3 (for a same non reduced basis). 
Until dimension $175$ the certificate is very likely to succeed
since the maximum certified relative error is small.    
On several tenths of trials, the certificate never failed, with 
a certified $\max |\mu_{ij}|$ as close to $1/2$ (with $\eta = 1/2$) 
as $0.4999916$. 
\begin{center}
\begin{minipage}{\textwidth}\begin{center}
{\footnotesize
$$
\begin{array}{|c|c|c|c|c|c|} \hline 
\text{\rm Dimension} &  75 & 100 & 125 & 150 & 175  \\  \hline 
(\delta,\eta)=(3/4,1/2), ~~\kappa _{\infty}(A) &  6 \times 10^5 & 5.2 \times 10^6 &
2.3 \times 10^8 & 1.3 \times 10^{10} & 2 \times 10^{11} \\ 
t_k-t=\min_i \{t_{i+1}-t\} \text{\rm ~in~(\ref{eq-proglov})}& 
1.3 \times 10^{37}&8.5 \times 10^{26}& 4.2 \times 10^{20}&3  \times 10^{15}& 1.2 \times 10^{12}\\
\text{\rm Max.~certified~relative~error~on~} |r_{ij}|
& 1.3 \times 10^{-9}& 3.4 \times 10^{-8} & 2.2
\times 10^{-6}& 2.1 \times 10^{-5} & 6.3 \times 10^{-3} \\\hline
(\delta,\eta)=(0.99,0.5001), ~~\kappa _{\infty}(A) &  2.4\times 10^4 & 4.6 \times 10^5 &
4 \times 10^5 & 4 \times 10^{7} & 9 \times 10^{8} \\ 
\text{\rm Max.~certified~relative~error~on~} |r_{ij}|
& 5.1 \times 10^{-10}& 2.5 \times 10^{-9} & 3.9 
\times 10^{-8}& 6 \times 10^{-7} & 9.5 \times 10^{-6} \\\hline
\end{array}
$$
Figure 7.3: Reducedness certificate output on ``random'' reduced
bases from knapsack problems,\\ $\max |a_{ij}|$ goes from $10^{45}$ ($n=50$)
down to $10^{25}$ ($n=175$).
}\end{center}
\end{minipage}
\end{center}

Beyond dimension $175$ with this type of reduced basis, the certificate 
starts to fail more often. On dimension $200$ with a conditioning about
$10^{12}$
with $(3/4,1/2)$, the error bound on the relative error approaches~$1$. 
The properness with $\eta=1/2$ may become impossible to check, and
ask  for a certificate with  $\eta=1/2 + \epsilon$, say
$\eta=0.5001$. Note that the Lov\'asz test~(\ref{eq-proglov})
seems to fail later thanks to much better error bounds on the
diagonal in general. 
On dimension $300$ for  $(3/4,1/2)$ the quality of the reduced bases
is too deteriorated ($\kappa _{\infty} \approx 10^{19}$), 
and the error bounding algorithm  fails with the impossibility 
of having a small spectral radius in~(\ref{subsec-boundrhoG}).
Nevertheless, on a typical example of dimension $300$ with 
a $(0.99,0.5001)$ reduced basis, 
the error bounding algorithm remains effective ($\kappa _{\infty} \approx
2.5 \times 10^{13}$, $\|H\|_{\infty} \approx 0.6$). 
The certificate may not be able to certify the actual reducedness of
the basis, for example with $\min_i \{t_{i}-t\} \approx - 4.12 \times 10^8$, and
a too
big absolute error bound $4.42 \times 10^8$. 
By changing the certificate parameters 
to $(\delta - \epsilon_1, \eta + \epsilon_2)=(0.985,0.515)$, 
the certificate succeeds again, and therefore is still able to certify a 
relevant information on the basis. 

The numerical limitations of the certificate are close to  
those identified in~\cite[Heuristic~4]{NgSt06} for the reduction
process itself. Indeed, on the knapsack bases,
it is claimed in~\cite{NgSt06} that 
a precision $n/4 +o(n)$ should suffice when using the floating point
reduction of~\cite{NgSt05}.  This means that $n\approx 200$ 
is a barrier with a $53$~bits precision ($64$~bits numbers). 
The eventuality of a link between
both limitations deserves to be further investigated.


\section{Conclusions} \label{sec-conclusion}

Between numerical approximation and computer algebra, we propose a
certificate for an (exact) algebraic/geometric property---the LLL
reducedness of a lattice basis.
This work, based on the fast computation of certified error bounds,
inherits from the verification methods approach. In particular,  
thanks to the IEEE arithmetic standard, 
the floating point errors do not put a curb on the objective of
certification. They may rather be mastered
and used for accelerating the programs. 
In error bound computation and property certification, 
the foreground of our study is to understand the 
compromize between the cost and the quality/effectiveness of 
bounds and certificates. In our case for instance, may we hope for an $O(n^2)$ effective
certificate? Various 
computer arithmetics come in the background, where floating
point computation, multi-precision, verification identities, midpoint-radius
intervals, and exact computation are collaborative tools. 

We think that our study raises several  
directions that deserve further investigations. 
The error bounding problem for the $R$ factor, and its finite
precision implementation should be better understood and improved, ingredients such as
diagonal scaling and other approximate $QR$ 
factorizations may be introduced. The usefulness of taking into account the
algorithm used for computing $\widetilde{R}$ should be studied 
(in a more restrictive  verification approach). 
A more general question is to know 
whether reducedness could be certified without resorting to the $QR$
factorization?

To our knowledge, 
the minimum precision required for a proven LLL variants is 
$1.6n +o(n)$ with the L$^2$ algorithm of~\cite{NgSt05,NgSt06}
(for $\delta$ close to $1$ and $\eta$ close to $1/2$).
Our experiments show we may certify reducedness for dimensions 
much higher than this worst-case limit ($n_{\max} \leq 53/1.6 \approx 33$). The certificate is
therefore very effective for a use complementary to reduction
heuristics in dimension greater than $n_{\max}$ with double
precision.
Noticing the fact that the certificate 
is sensitive to the numerical properties of the input basis, it is
worth studying its extensions to reduction algorithms and
reducedness certificates with adaptative precision, and sensitive
to the numerical quality.    \\


\noindent
{\bf Acknowledgements.} 
We thank Damien Stehl\'e for fruitful discussions around the floating
point reduction algorithms and heuristics, and for his help in
testing reduced bases.


{\small 
\bibliographystyle{plain}

}

\end{document}